\documentclass[aps, pra, twocolumn, superscriptaddress, amsmath, showpacs, tightenlines, footinbib, longbibliography]{revtex4-1}
\usepackage[colorlinks, breaklinks, citecolor=blue, linkcolor=blue,  urlcolor={blue}]{hyperref}
\usepackage{breakurl}
\usepackage{amsfonts}
\usepackage{amsmath}
\usepackage{amssymb}
\usepackage{mathrsfs}
\usepackage{epsfig}
\usepackage{graphicx}
\usepackage{caption}
\usepackage{bm}
\usepackage{float}
\usepackage[table]{xcolor}
\usepackage{braket}
\usepackage{amsmath}
\usepackage{amsfonts}
\usepackage{amssymb}
\usepackage{graphicx} 
\usepackage{arydshln} 
\usepackage{textcase}
\usepackage{booktabs}

\usepackage{CJKutf8}

\usepackage{multirow}
\usepackage{mathrsfs}
\usepackage{dcolumn}
\usepackage{bm}
\usepackage{tensor}
\usepackage{color}
\usepackage{epsfig}
\usepackage{placeins}
\usepackage{textcomp}
\usepackage{bbold}
\usepackage{float}
\usepackage{verbatim}
\usepackage{times}
\usepackage{siunitx}
\usepackage{chemformula}
\usepackage{physics}
\usepackage{multirow}
\usepackage{makecell}
\usepackage{ragged2e}
\usepackage{subcaption}
\usepackage{booktabs}
\definecolor{pink}{rgb}{.6902,.18824,.37647}

\begin{document}
	
	\title{Quantum Weak Force Sensing With Squeezed Magnomechanics}
	\author{Qian Zhang}\thanks{Co-first authors with equal contribution}
	\affiliation{Key Laboratory of Low-Dimensional Quantum Structures and Quantum Control of Ministry of Education, Department of Physics and Synergetic Innovation Center for Quantum Effects and Applications, Hunan Normal University, Changsha 410081, China}
	\author{Jie Wang}\thanks{Co-first authors with equal contribution}
	\affiliation{Key Laboratory of Low-Dimensional Quantum Structures and Quantum Control of Ministry of Education, Department of Physics and Synergetic Innovation Center for Quantum  Effects and Applications, Hunan Normal University, Changsha 410081, China}
	\author{Tian-Xiang Lu}
	\affiliation{College of Physics and Electronic Information, Gannan Normal University, Ganzhou 341000, Jiangxi, China}
	\author{Franco Nori}
	\affiliation{Theoretical Quantum Physics Laboratory, Cluster for Pioneering Research, RIKEN, Wako-shi, Saitama 351-0198, Japan}
	\affiliation{Center for Quantum Computing, RIKEN, Wako-shi, Saitama 351-0198, Japan}
	\affiliation{Department of Physics, University of Michigan, Ann Arbor, Michigan 48109-1040, USA}
	\author{Hui Jing}
	\email{jinghui73@foxmail.com}
	\affiliation{Key Laboratory of Low-Dimensional Quantum Structures and Quantum Control of Ministry of Education, Department of Physics and Synergetic Innovation Center for Quantum  Effects and Applications, Hunan Normal University, Changsha 410081, China}
	\affiliation{Academy for Quantum Science and Technology, Zhengzhou University of Light Industry, Zhengzhou 450002, China}
	\date{\today}

\begin{abstract}
	Cavity magnomechanics, exhibiting remarkable experimental tunability, rich magnonic nonlinearities, and compatibility with various quantum systems, has witnessed considerable advances in recent years. However, the potential benefits of using cavity magnomechanical (CMM) systems in further improving the performance of quantum-enhanced sensing for weak forces remain largely unexplored. Here we show that the performance of a quantum CMM sensor can be significantly enhanced beyond the standard quantum limit (SQL), by squeezing the magnons. We find that, for comparable parameters, 2 orders of enhancement in force sensitivity can be achieved in comparison with the case without the magnon squeezing. Moreover, we show optimal parameter regimes of homodyne angle for minimizing added quantum noise. Our findings provide a promising approach for highly tunable and compatible quantum force sensing using hybrid CMM devices, with potential applications ranging from quantum precision measurements to quantum information processing.
\end{abstract}

\maketitle

\section{INTRODUCTION}

Hybrid cavity magnomechanical (CMM) systems \cite{zhang2016_Sci.Adv.} based on collective spin excitations (i.e., magnons) in ferromagnetic crystals (e.g., yttrium iron garnet YIG) with remarkable experimental tunability and compatibility~\cite{zhang2014_Phys.Rev.Lett.,tabuchi2014_Phys.Rev.Lett.,lachance-quirion2019_Appl.Phys.Express,potts2021_Phys.Rev.X,zuo2024_}, enabling strong and coherent interaction between magnons and photons, phonons, as well as superconducting qubits, have become a versatile platform for fundamental studies with a wide range of applications, e.g., electromagnetically induced transparency and absorption~\cite{zhang2016_Sci.Adv.}, quantum entanglement~\cite{jie2023_Phys.Rev.Lett.,yu2020_Nature,fan2023_Laser&PhotonicsReviews,fan2023_Phys.Rev.A}, quantum control of a single magnon~\cite{xu2023_Phys.Rev.Lett.a}, nonreciprocal magnon blockade~\cite{wang2021_Sci.ChinaPhys.Mech.Astron.}, nonclassical states~\cite{li2019_NewJ.Phys.,li2023_Natl.Sci.Rev.,li2021_QuantumSci.Technol.}, magnonic frequency comb~\cite{xu2023_Phys.Rev.Lett.}, quantum parametric amplification \cite{wang2023_Sci.ChinaPhys.Mech.Astron.} and mechanical cooling~\cite{kani2022_Phys.Rev.Lett.}. In very recent experiment, broken PT-symmetry was achieved via Non-Hermiticity caused by on-site loss, generating complex-valued edge states \cite{jie2024}. Particularly, the YIG sphere, which is rich in Kerr-type nonlinearity due to magnetocrystalline anisotropy \cite{wang2016_Phys.Rev.B}, has emerged as a promising new tool with extensive applications, such as quantum entanglement~\cite{zhang2019_Phys.Rev.Research,yang2022_Phys.Rev.A}, bi- and multi-stabilities~\cite{zhang2019_Sci.ChinaPhys.Mech.Astron.,wang2018_Phys.Rev.Lett.,shen2021_Phys.Rev.Lett.,shen2022_Phys.Rev.Lett.a}, ground-state cooling~\cite{asjad2022_FundamentalResearch,Zoepfl_Phys.Rev.Lett.}, and nonreciprocal effects~\cite{kong2019_Phys.Rev.Applied}. 
\begin{figure*}[t]
	\centering
	\includegraphics[width=2.0\columnwidth]{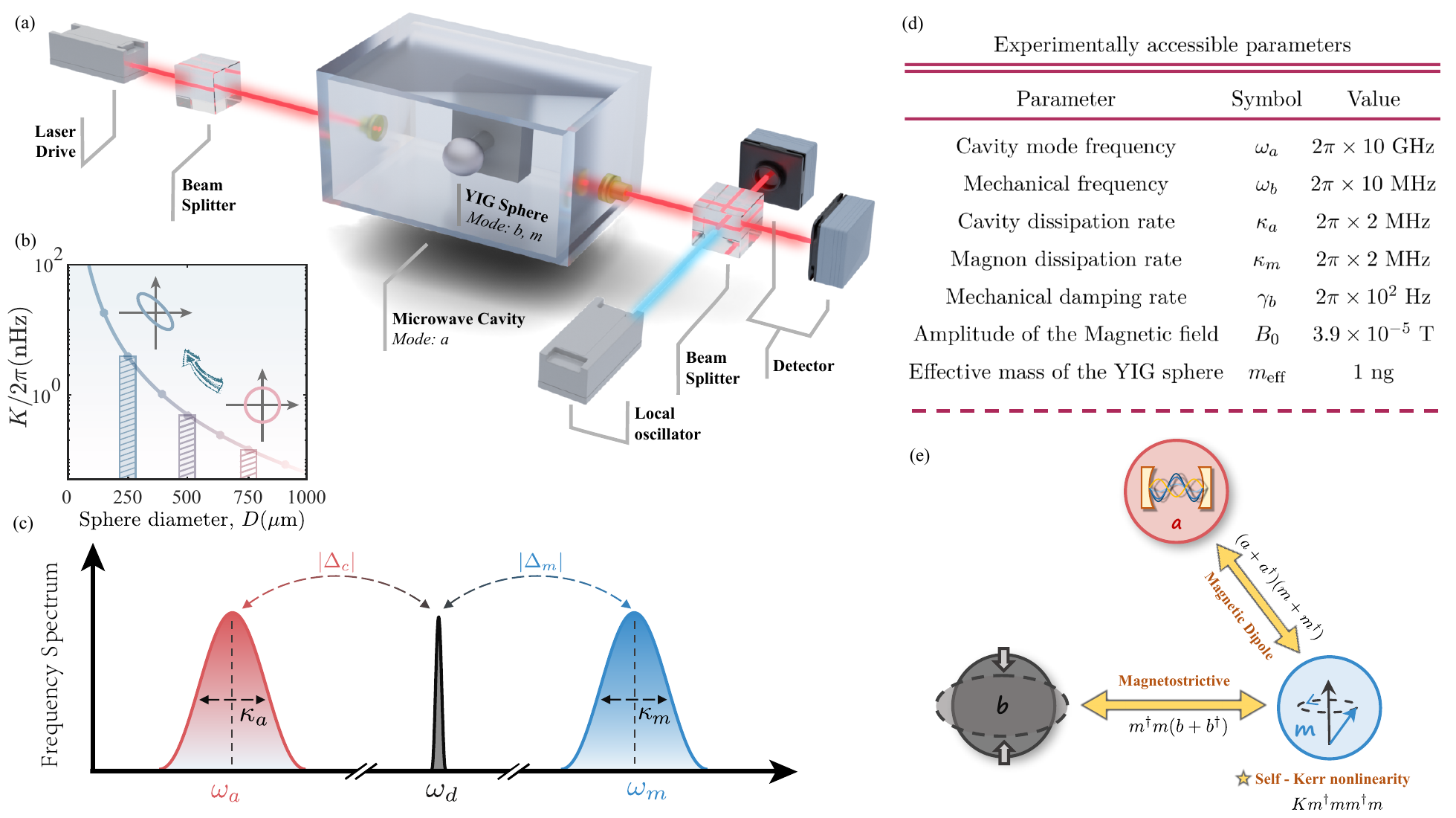}
	\caption{(a) Schematic of the hybrid CMM system. Magnon squeezing is achieved by a YIG sphere, which correlates the amplitude and phase fluctuations of the magnon mode. The output light field is measured with a balanced homodyne detector. (b) The magnon Kerr strength $K$ (log scale) is plotted as a function of the diameter $D$ of the YIG sphere \cite{zhang2019_Sci.ChinaPhys.Mech.Astron.}. As the diameter $D$ decreases, the magnon Kerr strength $K$ increases (i.e., $K$ is inversely proportional to $D$). (c) The frequency spectrum and linewidths of the system. The frequency of the drive field is blue detuned by $\Delta_{c}$ from the cavity resonance frequency $\omega_a$ and by $\Delta_m$ from the magnon resonance frequency. The lindwidth of the cavity mode and the magnon mode are $\kappa_{a}$ and $\kappa_{m}$, respectively. (d) Experimentally accessible parameters used in the numerical calculations. (e) Schematic diagram of the different interactions in a hybrid CMM system \cite{lachance-quirion2019_Appl.Phys.Express}, wherein the magnons couple to photons through magnetic dipole interaction as well as to phonons through magnetostrictive interaction. The optomechanical interaction is neglected given that the size of the YIG sphere ($0.1-1\,\mathrm{mm}$) is much smaller than the wavelength of the microwave field.} \label{fig1}
\end{figure*}

In parallel, quantum-enhanced sensing with the aid of quantum entanglement or squeezing has been widely used for the measurement of force~\cite{mason2019_Nat.Phys.,zhao2019_Sci.ChinaPhys.Mech.Astron.,Gebremariam2020_Sci.ChinaPhys.Mech.Astron.}, displacement~\cite{schliesser2008_NewJ.Phys.,hu2013_Front.Phys.}, electromagnetic fields~\cite{forstner2012_Phys.Rev.Lett.,forstner2014_Adv.Mater.,gotardo2023_Opt.Express}, acceleration~\cite{krause2012_NaturePhoton,cervantes2014_Appl.Phys.Lett.}, single spin~\cite{koppenhofer2023_Phys.Rev.Lett.}, mass~\cite{yu2016_NatCommun,sansa2020_NatCommun} due to their superior sensitivity to external perturbations. Recently, by using entanglement-enhanced joint force measurements with two optomechanical sensors, the SQL was surpassed scaling for arrayed optomechanical sensors \cite{xia2023_Nat.Photon.}. However, how to achieve sub-SQL quantum force sensing by utilizing the unique advantages of CMM systems has remained unexplored till now. 

Here in this work, we propose to build a bridge between such two active fields as quantum engineering of CMM systems and quantum-ehanced sensing. We show that the performance of a CMM sensor can be significantly enhanced drawing support from the magnon squeezing. We show that tuning the magnon Kerr strength can mitigate the  effects of backaction noise \cite{potts2021_Phys.Rev.X}, allowing for a considerable suppression of quantum noise, which is 2 orders of magnitude less (that is, much more sensitive) than the SQL. Our work is compatible with existing CMM techniques, and all parameters used for numerical calculations are experimentally accessible~\cite{zhang2014_Phys.Rev.Lett.,zhang2016_Sci.Adv.}.

\section{QUANTUM-ENHANCED FORCE SENSING BY SQUEEZING MAGNONS}

In recent experiments, the CMM devices with remarkable experimental tunability have been utilized to achieve the bi- and multi-stability~\cite{wang2018_Phys.Rev.Lett.,shen2022_Phys.Rev.Lett.}, the ternary logic gate and long-time memory~\cite{shen2021_Phys.Rev.Lett.}, magnomechanically induced transparency and absorption~\cite{zhang2016_Sci.Adv.}, to name a few~\cite{shen2021_Phys.Rev.Lett.}. We also note that advanced techniques based on non-classical squzzed light resources been developed to overcome the SQL for individual sensors~\cite{xia2023_Nat.Photon.,chalopin2018_NatCommun,lawrie2019_ACSPhotonics}. It is worth noting that, the merits of the CMM system in generating quantum squeezing have already confirmed in experiments~\cite{li2023_Natl.Sci.Rev.}. However, as far as we know, the possibility of enhancing weak force sensing by utilizing squeezed CMM system has not been explored. 

Here, we consider a hybrid CMM system, as sketched in Fig.~\ref{fig1} (a), in which a strong coupling between microwave photons and magnons generated by a ferromagnetic YIG sphere placed inside the cavity is achieved via the magnetic dipole interaction, which can be adjusted by varying the position of the YIG sphere inside the cavity or the angle between the bias magnetic field and the microwave magnetic field \cite{zhang2016_Sci.Adv.}. The magnetostrictive force couples magnons and phonons \cite{lachance-quirion2019_Appl.Phys.Express}. The optomechanical interaction is neglected given that the size of the YIG sphere ($0.1-1\,\mathrm{mm}$) is much smaller than the wavelength of the microwave field. In the frame rotating at the drive frequency $\omega_d$, the Hamiltonian of the system can be described as
\begin{equation}
\begin{aligned}
H =  &~\hbar\Delta_{c}a^{\dagger}a+\hbar\Delta_mm^{\dagger}m+\frac{\hbar}{2}\omega_{b}\left(Q^{2}+P^{2}\right)+\hbar g_{mb}m^{\dagger}mQ\\
& +\hbar g_{ma}\left(a^{\dagger}m+am^{\dagger}\right)+Km^{\dagger}mm^{\dagger}m+i\hbar\Omega\left(m^{\dagger}-m\right),
\end{aligned}
\end{equation}
where $a$ ($a^{\dag}$) and $m$ ($m^{\dag}$) represent the annihilation (creation) operator of the cavity mode (with frequency $ \omega_a$) and the magnon mode (with frequency $ \omega_m$), respectively. $\Delta_{c} (\Delta_m) =\omega_{a} (\omega_{m})-\omega_d$ denotes the detuning between the cavity (magnon) mode and the driving magnetic field. $Q$ and $P$ represent the dimensionless position and momentum quadratures of the mechanical mode at resonance frequency $\omega_b$, satisfying the commutation relation $[Q, P]\,{=}\, i$, $g_{ma}$ and $g_{mb}$ represent the magnon-photon coupling strength and the magnon-phonon coupling strength, respectively. $K=\mu_0K_{\rm an}\gamma ^2/(M^2V_m)$ is the strength of the magnon Kerr effect caused by the magnetocrystalline anisotropy, where $\mu_0$ is the vacuum permeability, $\gamma=2\pi\times28\,\mathrm{GHz/T}$ is the gyromagnetic ration, $M$ is the saturation magnetization, $K_{\rm an}$ and $V_m$ are the first-order anisotropy constant and the volume of the YIG sphere, respectively. Figure \ref{fig1}(b) shows $K$ is inversely proportional to the cube of diameter $D$ of the YIG sphere. We note that the diameter of the YIG sphere used in the experiment typically ranges from $0.1\,\mathrm{mm}$ to $1\,\mathrm{mm}$ , corresponding to $K$ ranges from $0.05\,\mathrm{nHz}$  to $100\,\mathrm{nHz}$~\cite{zhang2019_Sci.ChinaPhys.Mech.Astron.}. $\Omega=\sqrt{5}\gamma\sqrt{N}B_{0}/4$ represents strength of the drive field directly pumping the YIG sphere with the external magnetic field $B_{0}$, where $N=\rho V_m$ denotes the total number of spins with $\rho=4.22\times10^{27}\,{\rm m^{-3}}$ is the spin density of the YIG sphere.

It is worth noting that the CMM system was already well-established in experiments. For examples, a strongly coupled cavity-magnon system was utilized for realizing the bi- and multi-stability \cite{wang2018_Phys.Rev.Lett.,shen2022_Phys.Rev.Lett.}, and by selecting suitable input power, the ternary logic gate and long-time memory were experimentally demonstrated in a hybrid CMM system with the magnon Kerr effect \cite{shen2021_Phys.Rev.Lett.}. We also note that by introducing quantum squeezing or entanglement, an improved sensitivity in the shot-noise-dominant regime was achieved \cite{xiang2011_NaturePhoton}, and by tuning the pump phase of the optical parametric amplifier, considerably suppressed quantum noise and thus giant enhancement of force sensitivities were achieved \cite{zhao2019_Sci.ChinaPhys.Mech.Astron.}. Indeed, the merits of the CMM system in generating quantum squeezing have already confirmed in experiments \cite{li2023_Natl.Sci.Rev.}, and now our primary objective in this work is to investigate the impact of magnon squeezing on quantum-enhanced weak force sensing.
\begin{figure}[t]
	\centering
	\includegraphics[width=1.0\columnwidth]{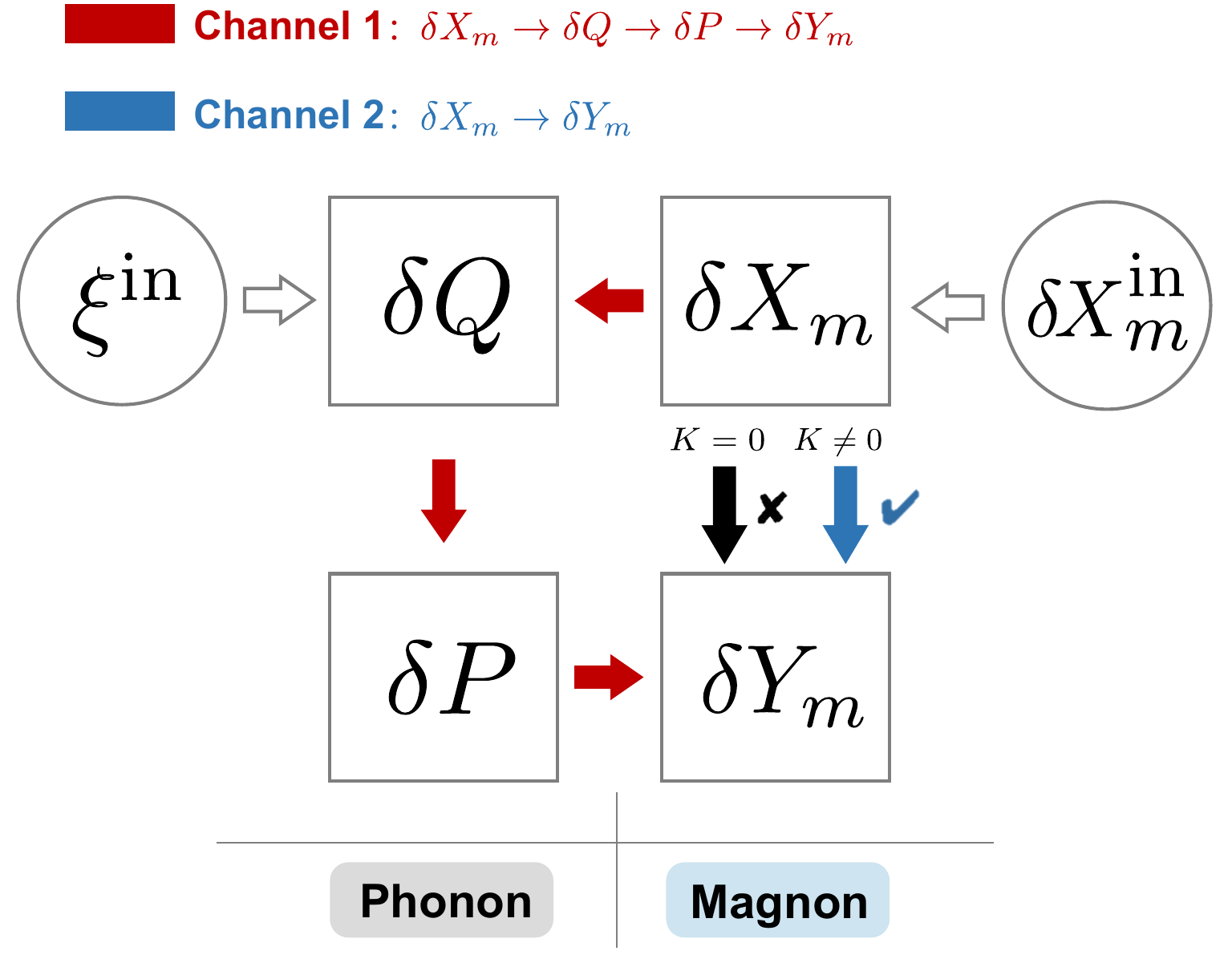}
	\caption{Flow chart of dynamical backaction noise. Channel 1 (red arrows) represents the original magnomechanical backaction noise due to magnetostrictive interaction. Another path (blue arrow) can be introducd by utilizing magnon Kerr effect. The destructive interference between Channel 1 and Channel 2 can reduce the impacts of the backaction noise and thus enhance the sensitivity of the force sensing.}
	\label{fig2}
\end{figure}
\begin{figure*}[t]
	\centering
	\includegraphics[width=\textwidth]{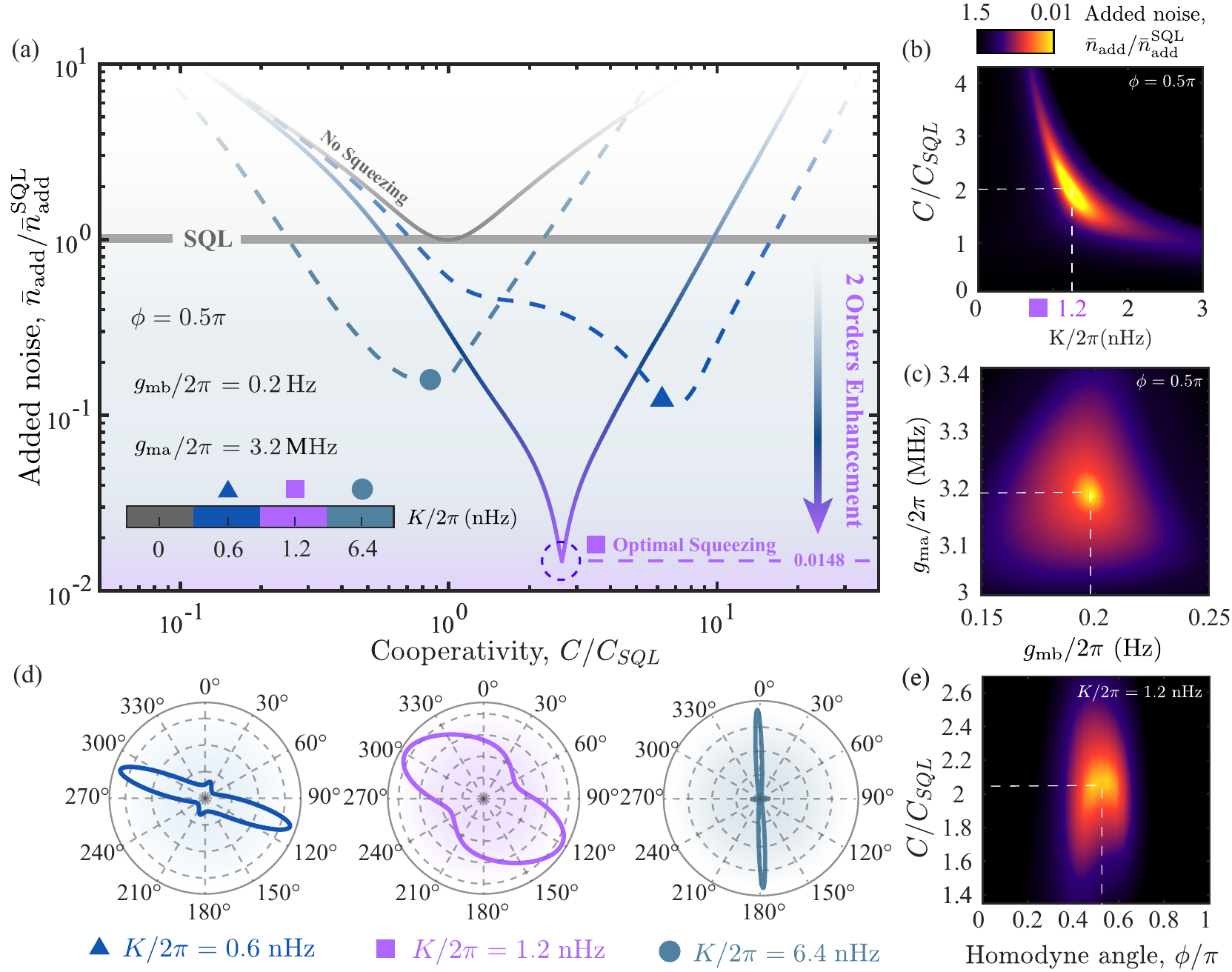}
	\captionsetup{justification=raggedright}
	\caption{(a) Weak force measurement below the standard quantum limit. Added Noise (log scale) is plotted as a function of the cooperativity $C/C_{\rm SQL}$ when $K/2\pi$ = 0.6, 1.2 and 6.4 nHz. The red solid line corresponds to the SQL. In the presence of magnon squeezing, measured added noise can be suppressed to 2 orders of magnitude less than the SQL. The magnon-photon coupling strength $g_{\rm ma}/2\pi=3.2$ MHz, and the magnomechanical coupling strength $g_{\rm mb}/2\pi=0.2$ Hz.
		(b) Added Noise $\bar{n}_{\rm add}/\bar{n}_{\rm add}^{\text{SQL}}$ versus the homodyne angle $\phi$ when $K/2\pi$ = 0.6, 1.2 and 6.4 nHz. Added Noise $\bar{n}_{\rm add}/\bar{n}_{\rm add}^{\rm SQL}$ versus (c) the magnon Kerr strength $K$ and the magnomechanical cooperativity $C/C_{\rm SQL}$. (d) the homodyne angle $\phi$ and the cooperativity $C/C_{\rm SQL}$ (e) the magnomechanical coupling strength $g_{\rm mb}$ and the magnon-photon coupling strength $g_{\rm ma}$. The experimental accessible parameters used for numerical calculations in our work is listed in Fig.~\ref{fig1}(d).}
	\label{fig3}
\end{figure*}

The quantum Langevin equations (valid when $\omega_a, \omega_m \gg g_{ma}, \kappa_{a}, \kappa_{m}$) describing the system are given by
\begin{equation}
\begin{aligned}
\dot{a}~=&-\left[i\Delta_{c}+\frac{\kappa_{a}}{2}\right]a-ig_{ma}m+\sqrt{\kappa_{a}}a^{{\rm in}},\\
\dot{m}=&-\left[i\Delta_m+\frac{\kappa_{m}}{2}\right]m-ig_{mb}mQ-ig_{ma}a\\
&-2iK m^{\dagger}mm+\Omega+\sqrt{\kappa_{m}}m^{{\rm in}},\\
\dot{Q}=&~\omega_{b}P,~~~\dot{P}=-\omega_{b}Q-g_{mb}m^{\dagger}m-\gamma_{b}P+\xi^{{\rm in}},
\end{aligned}
\end{equation}
where $\kappa_a$ ($a^{\rm in}$), $\kappa_m$ ($m^{\rm in}$), and $\gamma_{b}$ ($\xi^{\rm in}$) are the dissipation rate (input noise operators) of the cavity, magnon and mechanical modes, respectively. The steady-state mean values are
\begin{equation}
	\begin{aligned}
		a_{s}=&~\frac{-ig_{ma}\Omega}{\left(i\tilde{\Delta}_{m}+\frac{\kappa_{m}}{2}\right)\left(i\Delta_{c}+\frac{\kappa_{a}}{2}\right)+g_{ma}^{2}},\\
		m_{s}=&~\frac{\Omega\left(i\Delta_{c}+\frac{\kappa_{a}}{2}\right)}{\left(i\tilde{\Delta}_{m}+\frac{\kappa_{m}}{2}\right)\left(i\Delta_{c}+\frac{\kappa_{a}}{2}\right)+g_{ma}^{2}},\\
		Q_{s}=&-\frac{g_{mb}}{\omega_{b}}\left|m_{s}\right|^{2},
	\end{aligned}
\end{equation}
where $\tilde{\Delta}_m=\bar{\Delta}_m+2K\left|m_{s}\right|^{2}$ is the effective detuning between the magnon mode and the driving field with $\bar{\Delta}_m=\Delta_m+g_{mb} Q_s$. The noise forces acting on the mechanical membrane are $\xi^{{\rm in}}=\xi^{\mathrm{th}}+\xi^{\mathrm{ex}}$, where $\xi^{\mathrm{th}}$ and $\xi^{\mathrm{ex}}$ are the scaled thermal force and the detected force with dimension \si{\hertz^{\left.1\middle/2\right.}}, respectively. All noise operators are zero mean $\langle a^{\rm in}\rangle = \langle m^{\rm in} \rangle = \langle \xi^{\rm in} \rangle = 0$.

Considering the situation that a strong microwave drive field is applied to directly pump the YIG sphere, thus we can expand each operator as the sum of its classical mean value and a small quantum fluctuation~\cite{aspelmeyer2014_Rev.Mod.Phys.}, i.e., $\hat{a}=a_s+\delta\hat a$, $\hat{Q}=Q_s+\delta\hat{Q}$, and $\hat{P}=P_s+\delta\hat{P}$, with $\langle \delta \hat a \rangle = \langle \delta \hat Q\rangle = \langle \delta \hat P\rangle = 0$. By defining the quadrature fluctuation operators $\delta X_{a}=\left(\delta a+\delta a^{\dagger}\right)/\sqrt{2},\delta X_{m}=\left(\delta m+\delta m^{\dagger}\right)/\sqrt{2},\delta Y_{a}=\left(\delta a-\delta a^{\dagger}\right)/\sqrt{2}i,\delta Y_{m}=\left(\delta m-\delta m^{\dagger}\right)/\sqrt{2}i$ with the associated input noise operators $(\delta X_{k}^{{\rm in}}, \delta Y_{k}^{{\rm in}})(k=a,m)$ and corresponding correlation functions 
\begin{equation}
	\begin{aligned}
		\left\langle \delta X_{a(m)}^{\rm in}[\omega]\delta X_{a(m)}^{\rm in}[\omega^{\prime}]\right\rangle &=\frac{1}{2}\delta\left(\omega+\omega^{\prime}\right),\\\left\langle \delta Y_{a(m)}^{\rm in}[\omega]\delta Y_{a(m)}^{\rm in}[\omega^{\prime}]\right\rangle &=\frac{1}{2}\delta\left(\omega+\omega^{\prime}\right),\\\left\langle \delta X_{a(m)}^{\rm in}[\omega]\delta Y_{a(m)}^{\rm in}[\omega^{\prime}]\right\rangle &=\frac{i}{2}\delta\left(\omega+\omega^{\prime}\right),\\\left\langle \delta Y_{a(m)}^{\rm in}[\omega]\delta X_{a(m)}^{\rm in}[\omega^{\prime}]\right\rangle &=-\frac{i}{2}\delta\left(\omega+\omega^{\prime}\right),\\\left\langle \delta\xi^{\rm th}[\omega]\delta\xi^{\rm th}[\omega^{\prime}]\right\rangle &=\bar{n}_{m}^{T}\delta(\omega+\omega^{\prime}).
	\end{aligned}
\end{equation}
where $\bar{n}_{m}^{T}\approx\hbar\omega_{m}/k_{B}T$,  $k_{\mathrm{B}}$ is Boltzmann’s constant, and $T$ is the bath temperature. Then the linearized quantum Langevin equations describing the quadrature fluctuations $(\delta X_{a}, \delta Y_{a}, \delta X_{m}, \delta Y_{m}, \delta Q, \delta P)$ is given by
\begin{equation}
\begin{aligned}{\label{QLE}}
\delta\dot{X}_{a}=&-\frac{\kappa_{a}}{2}\delta X_{a}+\Delta_{c}\delta Y_{a}+g_{am}\delta Y_{m}+\sqrt{\kappa_{a}}\delta X_{a}^{\rm in},\\\delta\dot{Y}_{a}=&-\Delta_{c}\delta X_{a}-\frac{\kappa_{m}}{2}\delta Y_{a}-g_{am}\delta X_{m}+\sqrt{\kappa_{a}}\delta Y_{a}^{\rm in},\\\delta\dot{X}_{m}=&-\frac{\kappa_{m}}{2}\delta X_{m}+\tilde{\Delta}_{m}\delta Y_{m}+g_{am}\delta Y_{a}-G\delta Q+\sqrt{\kappa_{m}}\delta X_{m}^{\rm in},\\\delta\dot{Y}_{m}=&-\tilde{\Delta}_{m}\delta X_{m}-\frac{\kappa_{m}}{2}\delta Y_{m}-g_{am}\delta X_{a}+\sqrt{\kappa_{m}}\delta Y_{m}^{\rm in},\\\delta\dot{Q}=&~\omega_{m}\delta P,~~\delta\dot{P}=~G\delta Y_{m}-\omega_{m}\delta Q-\gamma_{b}\delta P+\xi^{{\rm in}},
\end{aligned}
\end{equation}
where $G=i\sqrt{2}g_{mb}m_{s}$ is the effective magnomechanical coupling strength.

We use a flow chart to depict the flow of backaction noise, as shown in Fig.~\ref{fig2}. For a conventional CMM system, magnons impart a radiation-pressure-like force on the phonons due to magnetostrictive interaction, introducing the magnomechanical backaction noise (Channel 1, red arrows in Fig.~\ref{fig2}).  In the presence of the magnon Kerr effect, a direct coupling between magnon quadratures emerges, introducing a magnon-induced backaction noise path (Channel 2, blue arrow)~\cite{potts2021_Phys.Rev.X}. As Fig.~\ref{fig2} shows, the destructive interference between the two channels suppresses the backaction noise of the CMM sensor and thus enhances the sensitivity of the force sensing.

In practice, one often measures the field that escapes from the cavity to extract information about the intracavity field because directly measuring the intracavity field is typically challenging. Thus, with the help of the input–output relation $\delta a^{\text{out}}=\sqrt{\kappa_{a}}\delta a-\delta a^{\text{in}}$, one can read out the frequency-dependent force noise via homodyne detection. The output field is mixed with a reference field (i.e., a local oscillator with phase $\phi$) at a $50$:$50$ beam splitter. Then the measured quadrature is given by:
\begin{equation}
\begin{aligned}
\delta X^{\text{out }}_{\phi,a}[\omega]=\delta X_{a}^{\text{out}}[\omega]\cos\phi+\delta Y_{a}^{\text{out}}[\omega]\sin\phi.
\end{aligned}
\end{equation}

The spectrum of the output field thus consists of amplitude or phase vacuum noise, thermal occupations, and quantum correlations (see Appendix for more details on definitions and calculations)
\begin{equation}
\begin{aligned}
\bar{S}_{\mathrm{II}}[\omega]&=\frac{1}{2}\left\langle \left\{ \delta X_{\phi,a}^{\text{out }}[\omega],\delta X_{\phi,a}^{\text{out }}[-\Omega]\right\} \right\rangle \\&=\mathcal{R}_{m}^{\phi}\left(\bar{n}_{m}+\bar{n}_{\rm add}[\omega]\right),
\end{aligned}
\end{equation}
where $\mathcal{R}_{m}^{\phi}$ is the mechanical response of our CMM sensor to the detected external force. Here, the added noise $\bar{n}_{\rm add}$ includes both imprecision noise and quantum backaction noise, contributing to the total force noise spectrum, which is essential for quantifying the sensitivity of force measurement
\begin{align}
\label{noise}
\bar{S}_{\mathrm{FF}}[\omega]=2\hbar m_{\mathrm{eff}}\kappa_{m}\omega_{b}(\bar{n}_{m}^{T}+\bar{n}_{{\rm add}}[\omega]).
\end{align}

In the following, we subtract the contributions of thermal noise to reveal the influence of the magnon squeezing on the quantum noise and the performance of the CMM sensor.

Quantum squeezing is known to be capable of increasing the sensitivity of quantum sensors~\cite{yu2020_Nature, xia2023_Nat.Photon.,xiang2011_NaturePhoton,martinciurana2017_Phys.Rev.Lett.,pezze2021_Nat.Photonics,brady2023_CommunPhys}. In our CMM system, magnon quadratures can be indirectly correlated through the mechanical mode via the magnetostrictive interaction. This correlation results in a quadrature squeezing of the magnon mode. And the magnon squeezing can also be achieved by utilizing the magnon Kerr nonlinearity.  As detailed in Ref.~\cite{li2023_Natl.Sci.Rev.}, the squeezing of the magnon mode can be transferred to the microwave cavity field, and the squeezing can be observed in the cavity output field via a homodyne detection. With identical parameters, the added noise is limited by the SQL for the non-squeezing case, while it can be suppressed below the SQL by tuning the magnon Kerr strength $K$.

As shown in Fig.~\ref{fig3}(a), the scaled added noise $\bar{n}_{\mathrm{add}}/\bar{n}_{\mathrm{add}}^{\mathrm{SQL}}$ is plotted for magnomechanical cooperativity $C=G^2/\kappa_m\gamma_b$ with different magnon Kerr strength $K$. For $K=0$ (i.e., in the absence of the magnon Kerr effect), there is always $\bar{n}_{\rm add}/\bar{n}_{\rm add}^{\rm SQL}>1$, in sharp contrast to the situation of $K>0$, for which $\bar{n}_{\rm add}/\bar{n}_{\rm add}^{\rm SQL}$ first decreases to below 1 (i.e., beating the SQL, the  the best sensitivity existing in the absence of quantum correlations, represented by the gray line in the Fig. \ref{fig3}(a)) with $C/C_{\rm SQL}$, and then increases with increasing $C/C_{\rm SQL}$ after reaching a minimum value. According to Eq.\,(\ref{noise}), the enhancement of the sensitivity induced by suppressed added noise is predicted. For $K/2\pi=\SI{1.2}{\nano\hertz}$, the added quantum noise can be suppressed by up to 2 orders of magnitude below the SQL, and thus significantly enhancing force sensing. 

To determine the parameters corresponding to the minimum noise, we plot $\bar{n}_{\mathrm{add}}/\bar{n}_{\mathrm{add}}^{\mathrm{SQL}}$ as a function of magnomechanical cooperativity $C$ and magnon Kerr strength $K$ or the magnon-phonon coupling strength $g_{\rm mb}$ and the magnon-photon coupling strength $g_{\rm ma}$ in Fig.\,\ref{fig3} (b-c). We can find that the minimum added noise occurs for $C/C_{\rm SQL}=2$, $K/2\pi=\SI{1.2}{\nano\hertz}$, $g_{\rm mb}/2\pi=\SI{0.2}{\hertz}$, and $g_{\rm ma}/2\pi=\SI{3.2}{\mega\hertz}$. Moreover, the 2 orders of magnitude improvement in sensitivity compared to the result in the absence of magnon squeezing, as depicted in Fig.~\ref{fig3}(a), can be further enhanced by optimizing the homodyne detection (see Fig.~\ref{fig3}(d-e)). Physically, these phenomena can be understood by the generation of magnon squeezing, which can be facilitated by the destructive interference between the two channels of the backaction noise (see the discussion of Fig.~\ref{fig2}).

\section{CONCLUSION AND OUTLOOK}

In summary, we study the impact of quantum squeezing induced by the quantum correlations between the magnon quadratures on quantum-enhanced weak force sensing. We show that the performance of the CMM sensors can be significantly enhanced with considerably suppressed added quantum noise, achieving 2 orders of magnitude less (that is, much more sensitive) than the SQL. Furthermore,  we also show the impact of the homodyne angle, magnon-phonon coupling strength, and magnon-photon coupling strength on the sensitivity of our CMM system. Our work not only reveals the significant impact of the magnon squeezing in the realm of quantum precision measurements, but also confirms that compact CMM systems can serve as powerful new tools for quantum sensing. we expect  that such tools can play unique roles in applications involving magnetic signals such as seismics,  geomagnetically-matching navigation, and biomedical dianosis. 

H.J. is supported by the NSFC (Grant No. 11935006), the Science and Technology Innovation Program of Hunan Province (Grant No. 2020RC4047), and the Hunan Provincial Major Sci-Tech Program (Grant No. 2023ZJ1010). T.-X.L. is supported by the NSFC (Grant No. 12205054), and Ph.D. Research Foundation (BSJJ202122). F.N. is supported in part by: Nippon Telegraph and Telephone Corporation (NTT) Research, the Japan Science and Technology Agency (JST) [via the Quantum Leap Flagship Program (Q-LEAP), and the Moonshot R\&D Grant Number JPMJMS2061], the Asian Office of Aerospace Research and Development (AOARD) (via Grant No. FA2386-201-4069), and the Office of Naval Research (ONR) (via Grant No. N62909-23-1-2074).

\begin{appendix}

\section{DERIVATION OF THE ADDED QUANTUM NOISE}\label{Appendix A}
\makeatletter
\renewcommand\theequation{A\@arabic\c@equation }
\makeatother 
\setcounter{equation}{0}
\setcounter{figure}{0}
The output fluctuations in the amplitude and phase quadratures of the cavity and magnon modes to be measured in the Fourier domain can be derived from
\begin{widetext}
	\begin{align}\label{matrix-coefficients}
			\begin{pmatrix}
				\delta X_{a}^{out}
				\\
				\delta Y_{a}^{out}
				\\
				\delta X_{m}^{out}
				\\
				\delta Y_{m}^{out}
			\end{pmatrix}=\begin{pmatrix}
				\mathcal{A}_{a}
				& \mathcal{B}_{a}
				& \mathcal{C}_{a}
				& \mathcal{D}_{a}
				& \mathcal{E}_{a}\\
				\mathcal{F}_{a}
				& \mathcal{G}_{a}
				& \mathcal{H}_{a}
				& \mathcal{I}_{a}
				& \mathcal{J}_{a}\\
				\mathcal{A}_{m}
				& \mathcal{B}_{m}
				& \mathcal{C}_{m}
				& \mathcal{D}_{m}
				& \mathcal{E}_{m}\\
				\mathcal{F}_{m}
				& \mathcal{G}_{m}
				& \mathcal{H}_{m}
				& \mathcal{I}_{m}
				& \mathcal{J}_{m}\\
			\end{pmatrix}\begin{pmatrix}
				\delta X_{a}^{\rm in}
				&\delta Y_{a}^{\rm in}
				&\delta X_{m}^{\rm in}
				&\delta Y_{m}^{\rm in}
				&\delta \xi^{\mathrm{in}}
			\end{pmatrix}^{\mathrm{T}},
	\end{align}
	\begin{align}
		\mathcal{A}_{a}&=\sqrt{\gamma}\mathbb{A}_{1}-1,&\mathcal{F}_{a}&=\sqrt{\gamma}\mathbb{A}_{3},&\mathcal{A}_{m}&=\sqrt{\gamma}\mathbb{A}_{2},&\mathcal{F}_{m}&=\sqrt{\gamma}\mathbb{A}_{4},\nonumber\\\mathcal{B}_{a}&=\sqrt{\gamma}\mathbb{B}_{1},&\mathcal{G}_{a}&=\sqrt{\gamma}\mathbb{B}_{3}-1,&\mathcal{B}_{m}&=\sqrt{\gamma}\mathbb{B}_{2},&\mathcal{G}_{m}&=\sqrt{\gamma}\mathbb{B}_{4},\nonumber\\\mathcal{C}_{a}&=\sqrt{\gamma}\mathbb{C}_{1},&\mathcal{H}_{a}&=\sqrt{\gamma}\mathbb{C}_{3},&\mathcal{C}_{m}&=\sqrt{\gamma}\mathbb{C}_{2}-1,&\mathcal{H}_{m}&=\sqrt{\gamma}\mathbb{C}_{4},\nonumber\\\mathcal{D}_{a}&=\sqrt{\gamma}\mathbb{D}_{1},&\mathcal{I}_{a}&=\sqrt{\gamma}\mathbb{D}_{3},&\mathcal{D}_{m}&=\sqrt{\gamma}\mathbb{D}_{2},&\mathcal{I}_{m}&=\sqrt{\gamma}\mathbb{D}_{4}-1,\nonumber\\\mathcal{E}_{a}&=\sqrt{\gamma}\mathbb{E}_{1},&J_{a}&=\sqrt{\gamma}\mathbb{E}_{3},&\mathcal{E}_{m}&=\sqrt{\gamma}\mathbb{E}_{2},&J_{m}&=\sqrt{\gamma}\mathbb{E}_{4}.
	\end{align}
\end{widetext}
\begin{widetext}
	\begin{align}
		\mathrm{A}_{1}&=\frac{\mathcal{E}_{+}\mathcal{B}_{-}+\mathcal{C}_{+}\mathcal{A}_{-}}{\mathcal{A}_{+}\mathcal{A}_{-}-\mathcal{B}_{+}\mathcal{B}_{-}}+\mu\left(\frac{\mathcal{A}_{-}\mathcal{H}_{+}+\mathcal{B}_{-}\mathcal{H}_{-}}{\mathcal{A}_{+}\mathcal{A}_{-}-\mathcal{B}_{+}\mathcal{B}_{-}}\right),&\mathrm{A}_{2}&=\frac{\mathcal{E}_{+}\mathcal{A}_{+}+\mathcal{C}_{+}\mathcal{B}_{+}}{\mathcal{A}_{+}\mathcal{A}_{-}-\mathcal{B}_{+}\mathcal{B}_{-}}+\mu\left(\frac{\mathcal{A}_{+}\mathcal{H}_{-}+\mathcal{B}_{+}\mathcal{H}_{+}}{\mathcal{A}_{+}\mathcal{A}_{-}-\mathcal{B}_{+}\mathcal{B}_{-}}\right),\nonumber\\\mathrm{B}_{1}&=\frac{\mathcal{F}_{-}\mathcal{B}_{-}+\mathcal{D}_{+}\mathcal{A}_{-}}{\mathcal{A}_{+}\mathcal{A}_{-}-\mathcal{B}_{+}\mathcal{B}_{-}}+\sigma\left(\frac{\mathcal{A}_{-}\mathcal{H}_{+}+\mathcal{B}_{-}\mathcal{H}_{-}}{\mathcal{A}_{+}\mathcal{A}_{-}-\mathcal{B}_{+}\mathcal{B}_{-}}\right),&\mathrm{B}_{2}&=\frac{\mathcal{F}_{-}\mathcal{A}_{+}+\mathcal{D}_{+}\mathcal{B}_{+}}{\mathcal{A}_{+}\mathcal{A}_{-}-\mathcal{B}_{+}\mathcal{B}_{-}}+\sigma\left(\frac{\mathcal{A}_{+}\mathcal{H}_{-}+\mathcal{B}_{+}\mathcal{H}_{+}}{\mathcal{A}_{+}\mathcal{A}_{-}-\mathcal{B}_{+}\mathcal{B}_{-}}\right),\nonumber\\\mathrm{C}_{1}&=\frac{\mathcal{E}_{-}\mathcal{A}_{-}+\mathcal{C}_{-}\mathcal{B}_{-}}{\mathcal{A}_{+}\mathcal{A}_{-}-\mathcal{B}_{+}\mathcal{B}_{-}}+\rho\left(\frac{\mathcal{A}_{-}\mathcal{H}_{+}+\mathcal{B}_{-}\mathcal{H}_{-}}{\mathcal{A}_{+}\mathcal{A}_{-}-\mathcal{B}_{+}\mathcal{B}_{-}}\right),&\mathrm{C}_{2}&=\frac{\mathcal{E}_{-}\mathcal{B}_{+}+\mathcal{C}_{-}\mathcal{A}_{+}}{\mathcal{A}_{+}\mathcal{A}_{-}-\mathcal{B}_{+}\mathcal{B}_{-}}+\rho\left(\frac{\mathcal{A}_{+}\mathcal{H}_{-}+\mathcal{B}_{+}\mathcal{H}_{+}}{\mathcal{A}_{+}\mathcal{A}_{-}-\mathcal{B}_{+}\mathcal{B}_{-}}\right),\nonumber\\\mathrm{D}_{1}&=\frac{\mathcal{F}_{+}\mathcal{A}_{-}+\mathcal{D}_{-}\mathcal{B}_{-}}{\mathcal{A}_{+}\mathcal{A}_{-}-\mathcal{B}_{+}\mathcal{B}_{-}}+\nu\left(\frac{\mathcal{A}_{-}\mathcal{H}_{+}+\mathcal{B}_{-}\mathcal{H}_{-}}{\mathcal{A}_{+}\mathcal{A}_{-}-\mathcal{B}_{+}\mathcal{B}_{-}}\right),&\mathrm{D}_{2}&=\frac{\mathcal{F}_{+}\mathcal{B}_{+}+\mathcal{D}_{-}\mathcal{A}_{+}}{\mathcal{A}_{+}\mathcal{A}_{-}-\mathcal{B}_{+}\mathcal{B}_{-}}+\nu\left(\frac{\mathcal{A}_{+}\mathcal{H}_{-}+\mathcal{B}_{+}\mathcal{H}_{+}}{\mathcal{A}_{+}\mathcal{A}_{-}-\mathcal{B}_{+}\mathcal{B}_{-}}\right),
	\end{align}
\end{widetext}
\begin{widetext}
	\begin{align}
		\eta&=\tilde{\Delta}_{m}g_{am}^{2}\chi_{0}^{3}\chi_{1}^{2}\chi\left(\frac{a_{1}q_{m}+a_{2}q_{a}}{a_{1}m_{1}-a_{2}m_{2}}\right)-g_{am}\chi_{0}\chi_{1}\chi\left(\frac{m_{1}q_{a}+m_{2}q_{m}}{a_{1}m_{1}-a_{2}m_{2}}\right)+\sqrt{2}Q_{m},\nonumber\\\mu&=G\tilde{\Delta}_{m}\varLambda\chi_{m}\chi_{2}\left(\frac{\mathcal{E}_{+}\mathcal{A}_{+}+\mathcal{C}_{+}\mathcal{B}_{+}}{\mathcal{A}_{+}\mathcal{A}_{-}-\mathcal{B}_{+}\mathcal{B}_{-}}\right)-G\varLambda_{2}\chi_{m}\chi_{2}\left(\frac{\mathcal{E}_{+}\mathcal{B}_{-}+\mathcal{C}_{+}\mathcal{A}_{-}}{\mathcal{A}_{+}\mathcal{A}_{-}-\mathcal{B}_{+}\mathcal{B}_{-}}\right)+\sqrt{\kappa_{a}}G\Delta_{a}\tilde{\Delta}_{m}g_{am}\chi_{0}^{4}\chi_{1}^{2}\chi\chi_{m}\chi_{2},\nonumber\\\sigma&=G\tilde{\Delta}_{m}\varLambda\chi_{m}\chi_{2}\left(\frac{\mathcal{F}_{-}\mathcal{A}_{+}+\mathcal{D}_{+}\mathcal{B}_{+}}{\mathcal{A}_{+}\mathcal{A}_{-}-\mathcal{B}_{+}\mathcal{B}_{-}}\right)-G\varLambda_{2}\chi_{m}\chi_{2}\left(\frac{\mathcal{F}_{-}\mathcal{B}_{-}+\mathcal{D}_{+}\mathcal{A}_{-}}{\mathcal{A}_{+}\mathcal{A}_{-}-\mathcal{B}_{+}\mathcal{B}_{-}}\right)-\sqrt{\kappa_{a}}G\tilde{\Delta}_{m}g_{am}\chi_{0}^{3}\chi_{1}^{2}\chi\chi_{m}\chi_{2},\nonumber\\\rho&=G\tilde{\Delta}_{m}\varLambda\chi_{m}\chi_{2}\left(\frac{\mathcal{E}_{-}\mathcal{B}_{+}+\mathcal{C}_{-}\mathcal{A}_{+}}{\mathcal{A}_{+}\mathcal{A}_{-}-\mathcal{B}_{+}\mathcal{B}_{-}}\right)-G\varLambda_{2}\chi_{m}\chi_{2}\left(\frac{\mathcal{E}_{-}\mathcal{A}_{-}+\mathcal{C}_{-}\mathcal{B}_{-}}{\mathcal{A}_{+}\mathcal{A}_{-}-\mathcal{B}_{+}\mathcal{B}_{-}}\right)-\sqrt{\kappa_{m}}G\tilde{\Delta}_{m}\chi_{0}^{2}\chi_{1}\chi\chi_{m}\chi_{2},\nonumber\\\nu&=G\tilde{\Delta}_{m}\varLambda\chi_{m}\chi_{2}\left(\frac{\mathcal{F}_{+}\mathcal{B}_{+}+\mathcal{D}_{-}\mathcal{A}_{+}}{\mathcal{A}_{+}\mathcal{A}_{-}-\mathcal{B}_{+}\mathcal{B}_{-}}\right)-G\varLambda_{2}\chi_{m}\chi_{2}\left(\frac{\mathcal{F}_{+}\mathcal{A}_{-}+\mathcal{D}_{-}\mathcal{B}_{-}}{\mathcal{A}_{+}\mathcal{A}_{-}-\mathcal{B}_{+}\mathcal{B}_{-}}\right)+\sqrt{\kappa_{m}}G\chi_{0}\chi_{1}\chi\chi_{m}\chi_{2},
	\end{align}
\end{widetext}
\begin{equation}
	\begin{aligned}
		\mathrm{A}_{3}&=\Delta_{a}\varLambda_{1}\mathrm{A}_{1}-\varLambda_{2}\mathrm{A}_{3}+\sqrt{2}Q_{a}\mu-\sqrt{\kappa_{a}}\Delta_{a}\chi_{0}^{2}\chi_{1}\chi,\\\mathrm{B}_{3}&=\Delta_{a}\varLambda_{1}\mathrm{B}_{1}-\varLambda_{2}\mathrm{B}_{3}+\sqrt{2}Q_{a}\sigma+\sqrt{\kappa_{a}}\chi_{0}\chi_{1}\chi,\\\mathrm{C}_{3}&=\Delta_{a}\varLambda_{1}\mathrm{C}_{1}-\varLambda_{2}\mathrm{C}_{3}+\sqrt{2}Q_{a}\rho+\sqrt{\kappa_{m}}\Delta_{a}\tilde{\Delta}_{m}g_{am}\chi_{0}^{4}\chi_{1}^{2}\chi,\\\mathrm{D}_{3}&=\Delta_{a}\varLambda_{1}\mathrm{D}_{1}-\varLambda_{2}\mathrm{D}_{3}+\sqrt{2}Q_{a}\nu-\sqrt{\kappa_{m}}\Delta_{a}g_{am}\chi_{0}^{3}\chi_{1}^{2}\chi,\\\mathrm{A}_{4}&=\tilde{\Delta}_{m}\varLambda_{1}\mathrm{A}_{3}-\varLambda_{2}\mathrm{A}_{1}+\sqrt{2}Q_{m}\mu+\sqrt{\kappa_{a}}\Delta_{a}\tilde{\Delta}_{m}g_{am}\chi_{0}^{4}\chi_{1}^{2}\chi,\\\mathrm{B}_{4}&=\tilde{\Delta}_{m}\varLambda_{1}\mathrm{B}_{3}-\varLambda_{2}\mathrm{B}_{1}+\sqrt{2}Q_{m}\sigma-\sqrt{\kappa_{a}}\tilde{\Delta}_{m}g_{am}\chi_{0}^{3}\chi_{1}^{2}\chi,\\\mathrm{C}_{4}&=\tilde{\Delta}_{m}\varLambda_{1}\mathrm{C}_{3}-\varLambda_{2}\mathrm{C}_{1}+\sqrt{2}Q_{m}\rho-\sqrt{\kappa_{a}}\tilde{\Delta}_{m}\chi_{0}^{2}\chi_{1}\chi,\\\mathrm{D}_{4}&=\tilde{\Delta}_{m}\varLambda_{1}\mathrm{D}_{3}-\varLambda_{2}\mathrm{D}_{1}+\sqrt{2}Q_{m}\nu+\sqrt{\kappa_{a}}\chi_{0}\chi_{1}\chi,
	\end{aligned}
\end{equation}
with
\begin{equation}
	\begin{aligned}
		\chi_{0}&=\left(-i\omega+\frac{\kappa_{m}}{2}\right)^{-1},\chi_{1}=\left(1+\Delta_{a}^{2}\chi_{0}^{2}\right)^{-1},\\\chi_{2}&=\left(1-G\chi_{m}\eta\right)^{-1},\chi_{3}=\left(1-\Delta_{a}\tilde{\Delta}_{m}g_{am}^{2}\chi_{0}^{4}\chi_{1}^{2}\right)^{-1},\\\varLambda_{1}&=g_{am}^{2}\chi_{0}^{3}\chi_{1}^{2}\chi,\quad\quad\varLambda_{2}=g_{am}\chi_{0}\chi_{1}\chi,\\\chi_{m}&=\omega_{m}\left(\omega_{m}^{2}-\omega^{2}-i\omega\gamma_{b}\right)^{-1},
	\end{aligned}
\end{equation}
\begin{figure}[b]
	\centering
	\includegraphics[scale=0.5]{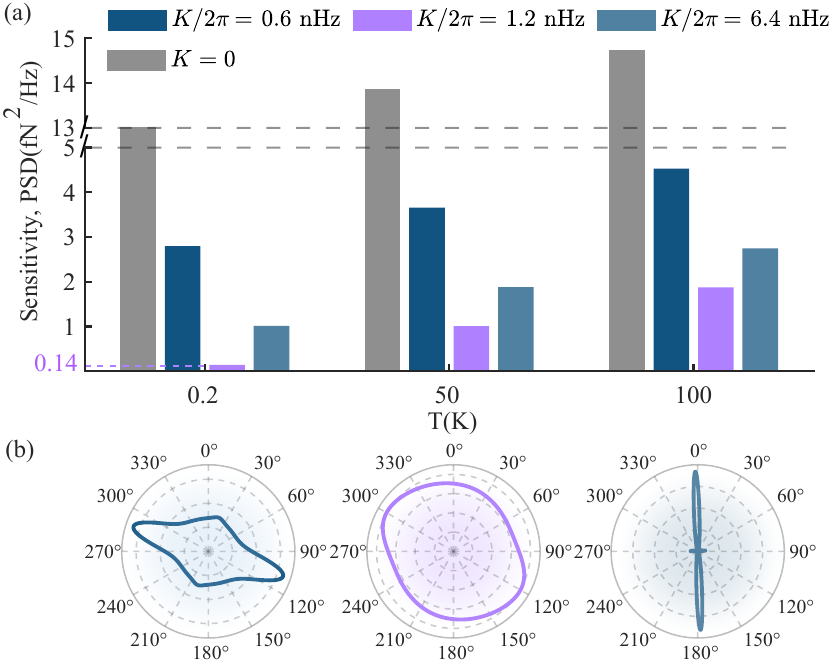}
	\caption{Noise power spectral density (PSD) of homodyne measurements versus (a) the homodyne angle $\phi$ (b) the temperature when $K/2\pi$ = 0.6, 1.2, 6.4 nHz. The parameters are the same as those in Fig.~\ref{fig3}. } \label{figS1}
\end{figure}
and
\begin{equation}
	\begin{aligned}
		\mathcal{A}_{\pm}&=1-\Delta_{\pm}^{2}g_{am}^{2}\chi_{0}^{4}\chi_{1}^{2}\chi+g_{am}^{2}\chi_{0}^{2}\chi_{1}\chi,\\\mathcal{B}_{\pm}&=\Delta_{\pm}g_{am}^{3}\chi_{0}^{4}\chi_{1}^{2}\chi-\Delta_{\mp}g_{am}\chi_{0}^{2}\chi_{1}\chi,\\\mathcal{C}_{\pm}&=\sqrt{\kappa_{a}}\chi_{0}+\sqrt{\kappa_{a}}\Delta_{+}\Delta_{-}g_{am}^{2}\chi_{0}^{5}\chi_{1}^{2}\chi-\sqrt{\kappa_{a}}\Delta_{\pm}^{2}\chi_{0}^{3}\chi_{1}\chi,\\\mathcal{D}_{\pm}&=\sqrt{\kappa_{a}}\Delta_{\pm}\chi_{0}^{2}\chi_{1}\chi-\sqrt{\kappa_{a}}\Delta_{\mp}g_{am}^{2}\chi_{0}^{4}\chi_{1}^{2}\chi,
	\end{aligned}
\end{equation}
\begin{equation}
	\begin{aligned}
		\mathcal{E}_{\pm}&=\sqrt{\kappa_{a}}\Delta_{\mp}^{2}\Delta_{\pm}g_{am}\chi_{0}^{5}\chi_{1}^{2}\chi-\sqrt{\kappa_{a}}\Delta_{\pm}g_{am}\chi_{0}^{3}\chi_{1}\chi,\\\mathcal{F}_{\pm}&=\sqrt{\kappa_{a}}g_{am}\chi_{0}^{2}\chi_{1}\chi-\sqrt{\kappa_{a}}\Delta_{\pm}^{2}g_{am}\chi_{0}^{4}\chi_{1}^{2}\chi,\\\mathcal{G}_{\pm}&=\Delta_{\pm}g_{am}g_{mb}m_{s}\chi_{0}^{3}\chi_{1}^{2}\chi,\\\mathcal{H}_{\pm}&=\sqrt{2}Q_{\pm}\Delta_{\pm}\chi_{0}+\sqrt{2}Q_{\mp}g_{am}\chi_{0},
	\end{aligned}
\end{equation}

We measure the force noise by homodyne detection, in which the photocurrent $I$ is proportional to the rotated field quadrature
\begin{equation}
	\begin{aligned}
		\delta X^{\text{out }}_{\phi,a}[\omega]=\delta X_{a}^{\text{out}}[\omega]\cos\phi+\delta Y_{a}^{\text{out}}[\omega]\sin\phi,
	\end{aligned}
\end{equation}
\begin{figure}[t]
	\centering
	\includegraphics[width=1.0\columnwidth]{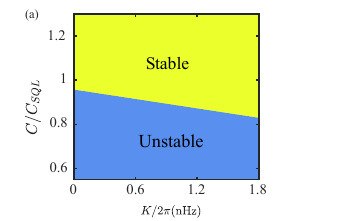}
	\caption{ The stable and unstable regions can be tuned by altering the magnon Kerr strength or the magnomechanical cooperativity, The parameters are the same as those in Fig.~\ref{fig2}. }\label{figS2}
\end{figure}
In the Fourier domain, the quadratures of the output fields to be measured are given by 
\begin{equation}
	\begin{aligned}
		\bar{S}_{\mathrm{II}}[\omega]&=\frac{1}{2}\left\langle \left\{ \delta X_{\phi,a}^{\text{out }}[\omega],\delta X_{\phi,a}^{\text{out }}[-\Omega]\right\} \right\rangle \\&=\mathcal{R}_{m}^{\phi}\left(\bar{n}_{m}+\bar{n}_{\rm add}[\omega]\right),
	\end{aligned}
\end{equation}
where the mechanical response of the system to the detected force signal is derived as
\begin{equation}
	\begin{aligned}
		\mathcal{R}_{m}^{\phi}=~&\mathcal{E}_{a}^{2}\cos^{2}\phi+\mathcal{J}_{a}^{2}\sin^{2}\phi+\mathcal{E}_{a}\mathcal{J}_{a}\sin\left(2\phi\right),
	\end{aligned}
\end{equation}
and the added noise consisting of the imprecision noise and backaction noise is
\begin{equation}
	\begin{aligned}
		\bar{n}_{{\rm add}}&=\frac{1}{2}\left(\frac{\mathcal{A}_{a}^{2}+\mathcal{B}_{a}^{2}+\mathcal{C}_{a}^{2}+\mathcal{D}_{a}^{2}+i\mathcal{A}_{a}\mathcal{C}_{a}+i\mathcal{B}_{a}\mathcal{D}_{a}}{\mathcal{E}_{a}^{2}\cos^{2}\phi+\mathcal{J}_{a}^{2}\sin^{2}\phi+\mathcal{E}_{a}\mathcal{J}_{a}\sin(2\phi)}\right)\cos^{2}\phi\\&+\frac{1}{2}\left(\frac{\mathcal{F}_{a}^{2}+\mathcal{G}_{a}^{2}+\mathcal{H}_{a}^{2}+\mathcal{I}_{a}^{2}+i\mathcal{F}_{a}\mathcal{H}_{a}+i\mathcal{G}_{a}\mathcal{I}_{a}}{\mathcal{E}_{a}^{2}\cos^{2}\phi+\mathcal{J}_{a}^{2}\sin^{2}\phi+\mathcal{E}_{a}\mathcal{J}_{a}\sin(2\phi)}\right)\sin^{2}\phi\\&+\frac{1}{4}\left(\frac{\mathcal{A}_{a}\mathcal{F}_{a}+\mathcal{B}_{a}\mathcal{G}_{a}+\mathcal{C}_{a}\mathcal{H}_{a}+\mathcal{D}_{a}\mathcal{I}_{a}}{\mathcal{E}_{a}^{2}\cos^{2}\phi+\mathcal{J}_{a}^{2}\sin^{2}\phi+\mathcal{E}_{a}\mathcal{J}_{a}\sin(2\phi)}\right)\sin(2\phi).
	\end{aligned}
\end{equation}

The sensitivity of force measurement is typically characterized by the noise power spectral density (PSD)
\begin{equation}
	\begin{aligned}
		\bar{S}_{\mathrm{FF}}[\omega]=2\hbar m_{\mathrm{eff}}\gamma_{b}\omega_{b}(\bar{n}_{m}^{T}+\bar{n}_{{\rm add}}).
	\end{aligned}
\end{equation}

In Fig. \ref{figS1}, similar features as in Fig. \ref{fig3}(a) arise, noise suppressions and thus giant enhancement of force sensitivities was achieved. Figure \ref{figS1} shows that the PSD of homodyne measurements shares the same dependency on the magnon Kerr strength $K$ as added quantum noise. Specifically, the PSD decreases as $K$ increases, until reaching a minimum value and then increases, which is in agreement with the performance of added quantum noise shown in Fig. \ref{fig2}.

\section{Stability analysis}\label{Appendix B}
\makeatletter
\renewcommand\theequation{B\@arabic\c@equation }
\makeatother \setcounter{equation}{0}
Finally, we analyze the stability properties of the CMM system by applying the Routh-Hurwitz criterion. 

The characteristic equation of $\left|\mathrm{C}-\lambda
\mathrm{I}\right|=0 $ can be reduced to:
\begin{equation}
a_{0}\lambda^{6}+a_{1}\lambda^{5}+a_{2}\lambda^{4}+a_{3}\lambda^{3}+a_{4}\lambda^{2}+a_{5}\lambda^{1}+a_{6}=0.
\end{equation}
where $a_{0}=1$, $a_{1}=2\kappa_{a}+\gamma_{b}$ and
\begin{equation}
\begin{aligned}
a_{2}=&\Delta_{c}^{2}+2g_{ma}^{2}+\frac{3\kappa_{a}^{2}}{2}+2\kappa_{a}\gamma_{b}+\tilde{\Delta}_m^{2}+\omega_{b}^{2},\\a_{3}=&\kappa_{a}\Delta_{c}^{2}+\gamma_{b}\left(\Delta_{c}^{2}+\frac{3\kappa_{m}^{2}}{2}+\tilde{\Delta}_m^{2}\right)+2g_{ma}^{2}\left(\kappa_{a}+\gamma_{b}\right)\\&+\sqrt{2}g_{mb}Gm_{s}\omega_{b}+\frac{\kappa_{a}^{3}}{2}+\kappa_{a}\tilde{\Delta}_m^{2}+2\kappa_{a}\omega_{b}^{2},
\end{aligned}
\end{equation}
\begin{equation}
\begin{aligned}
a_{4}=&\frac{1}{2}g_{ma}^{2}\left(-4\Delta_{c}\tilde{\Delta}_m+\kappa_{a}^{2}+4\kappa_{a}\gamma_{b}+4\omega_{b}^{2}\right)+\frac{1}{4}\kappa_{m}^{2}\\&+\gamma_{b}\left(\kappa_{a}\Delta_{c}^{2}+\frac{\kappa_{m}^{3}}{2}+\kappa\tilde{\Delta}_m^{2}\right)+\Delta_{c}^{2}\omega_{b}^{2}+\Delta_{c}^{2}\tilde{\Delta}_m^{2}\\&+g_{ma}^{4}+\frac{3\kappa_{a} g_{mb}Gm_{s}\omega_{b}}{\sqrt{2}}+\frac{\kappa_{m}^{4}}{16}+\frac{1}{4}\kappa_{a}^{2}\tilde{\Delta}_m^{2}+\tilde{\Delta}_m^{2},
\end{aligned}
\end{equation}
\begin{equation}
\begin{aligned}
a_{5}=&-2\Delta_{c}g_{ma}^{2}\gamma_{b}\tilde{\Delta}_m+\sqrt{2}\Delta_{c}^{2}g_{mb}Gm_{s}\omega_{b}+\frac{1}{4}\kappa_{a}^{2}\\&+\Delta_{c}^{2}\gamma_{b}\tilde{\Delta}_m^{2}+\kappa_{a}\Delta_{c}^{2}\omega_{b}^{2}+\sqrt{2}g_{ma}^{2}g_{mb}Gm_{s}\omega_{b}\\&+\frac{1}{2}\kappa_{a}^{2}g_{ma}^{2}\gamma_{b}+g_{ma}^{4}\gamma_{b}+2\kappa_{m} g_{ma}^{2}\omega_{b}^{2}+\frac{3\kappa_{m}^{2}g_{mb}}{2\sqrt{2}}\\&+\frac{1}{4}\kappa_{m}^{2}\gamma_{b}\tilde{\Delta}_m^{2}+\frac{\kappa_{m}^{4}\gamma_{b}}{16}+\kappa_{a}\tilde{\Delta}_m^{2}\omega_{b}^{2}+\frac{1}{2}\kappa_{a}^{3}\omega_{b}^{2},
\end{aligned}
\end{equation}
\begin{equation}
\begin{aligned}
a_{6}=&-2\Delta_{c}g_{ma}^{2}\tilde{\Delta}_m\omega_{b}^{2}+\frac{\kappa_{a}\Delta_{c}^{2}g_{mb}Gm_{s}\omega_{b}}{\sqrt{2}}+\frac{1}{4}\kappa_{a}^{2}\\&+\Delta_{c}^{2}\tilde{\Delta}_m^{2}\omega_{b}^{2}+\frac{\kappa_{m} g_{ma}^{2}g_{mb}Gm_{s}\omega_{b}}{\sqrt{2}}+\frac{1}{2}\kappa_{a}^{2}g_{ma}^{2}\omega_{b}^{2}\\&+g_{ma}^{4}\omega_{b}^{2}+\frac{\kappa_{m}^{3}g_{mb}Gm_{s}\omega_{b}}{4\sqrt{2}}+\frac{1}{4}\kappa_{m}^{2}\tilde{\Delta}_m^{2}\omega_{b}^{2}.
\end{aligned}
\end{equation}

By applying the criterion to the coefficient matrix $C$, we obtain the stability criterion of the system is defined by the inequalities
\begin{equation}
\begin{aligned}
\Delta_{n}=\left|\begin{array}{ccccc}
a_{1} & a_{0} & 0 & \vdots & 0\\
a_{3} & a_{2} & a_{1} & \vdots & 0\\
a_{5} & a_{4} & a_{3} & \vdots & 0\\
\cdots & \ldots & \ldots & \ldots & \ldots\\
0 & 0 & 0 & \vdots & a_{n}
\end{array}\right|>0
\end{aligned}
\end{equation}
where n = 1-6. The stable and unstable parameter regimes can be shown in the Fig.~\ref{figS2} and the parameters chosen in this paper are confirmed to be within the stable region.

\end{appendix}

\end{document}